\documentclass[fleqn,10pt]{wlscirep}
\usepackage[utf8]{inputenc}
\usepackage[T1]{fontenc}
\usepackage[nolist]{acronym}
\usepackage{nameref}

\newcommand{\etal}{\textit{ et al.}}

\title{Inter-Species Cell Detection: Datasets on pulmonary hemosiderophages in equine, human and feline specimens}

\author[1,2,+]{Christian~Marzahl}
\author[3]{Jenny Hill}
\author[3]{Jason Stayt}
\author[4]{Dorothee Bienzle}
\author[5]{Lutz Welker}
\author[1]{Frauke Wilm}
\author[2]{Jörn Voigt}
\author[6]{Marc Aubreville}
\author[1]{Andreas~Maier}
\author[7]{Robert~Klopfleisch}
\author[1,8]{Katharina Breininger}
\author[7,9]{Christof~A.~Bertram}

\affil[1]{Pattern Recognition Lab, Computer Science\\Friedrich-Alexander-Universit{\"a}t Erlangen-N{\"u}rnberg, Germany}
\affil[2]{Research and Development, EUROIMMUN Medizinische Labordiagnostika AG, Lübeck, Germany}
\affil[3]{VetPath Laboratory Services, Ascot, Western, Australia}
\affil[4]{Department of Pathobiology, OntarioVeterinary College, University of Guelph, Guelph, ON, Canada}
\affil[5]{Cytology Laboratory, Lungen Clinic Grosshansdorf, Germany}
\affil[6]{Technische Hochschule Ingolstadt, Ingolstadt, Germany}
\affil[7]{Institute of Veterinary Pathology, Freie Universit{\"a}t Berlin, Germany}
\affil[8]{Department of Artifical Intelligence in Biomedical Engineering, Friedrich-Alexander-Universität Erlangen-Nürnberg}
\affil[9]{Institute of Pathology, University of Veterinary Medicine, Vienna, Austria}
\affil[+]{corresponding author}

\begin{abstract}

\ac{phem} occurs among multiple species and can have various causes. 
Cytology of \ac{balf} using a 5-tier scoring system of alveolar macrophages based on their hemosiderin content is considered the most sensitive diagnostic method.
We introduce a novel, fully annotated multi-species \ac{phem} dataset which consists of 74 cytology \acp{wsi} with equine, feline and human samples. To create this high-quality and high-quantity dataset, we developed an annotation pipeline combining human expertise with deep learning and data visualisation techniques. We applied a deep learning-based object detection approach trained on 17 expertly annotated equine \acp{wsi}, to the remaining 39 equine, 12 human and 7 feline \acp{wsi}.
The resulting annotations were semi-automatically screened for errors on multiple types of specialised annotation maps and finally reviewed by a trained pathologists. Our dataset contains a total of 297,383 hemosiderophages classified into five grades. It is one of the largest publicly available \acp{wsi} datasets with respect to the number of annotations, the scanned area and the number of species covered. 

\end{abstract}
\begin{document}

\flushbottom
\maketitle

\thispagestyle{empty}

\acresetall

\section{Background \& Summary}

In recent years, deep learning has revolutionised microscopy-based image recognition. Outstanding results can be achieved in well-defined tasks under the condition that sufficient high-quality datasets are available~\cite{hou2020dataset,aubreville2020completely,bertram2019large}. For certain species and/or certain pathologies, however, available data may be sparse. Approaches such as transfer learning and domain adaptation provide the possibility to develop algorithms that generalise across species although they come with their own challenges and limitations~\cite{aubreville2020completely}. The generalised applicability of deep learning models between species could offer enormous scientific and economic value. For domains that lack appropriate training data, for example due to data protection and privacy restrictions, approaches that allow for this transferability may especially be useful in the context of animal models for human diseases.

To be able to develop, investigate and apply these algorithms, suitable cross-species datasets have to be available. The dataset described in this work aims to tackle several gaps present in currently available datasets.
Firstly, whereas there are a couple of highly domain and target specific \ac{wsi} datasets publicly available for tissue~\cite{hou2020dataset,aubreville2020completely,bertram2019large}, to the authors' knowledge none for cytologic research questions.
Secondly, no publicly available dataset provides annotated \acp{wsi} from multiple species for the same pathology. 
Finally, as shown in our previous publication~\cite{marzahl2020deep,10.1007/978-3-030-59710-8_3}, there is a high inter- and intra-observer variability for grading pulmonary hemosiderophages, which can be reduced by algorithmically supporting experts during labelling. For the development of these algorithms, large high-quality datasets are required, which is a further motivation for this publication. 

In the following sections, we will describe the creation of a novel multi-species \ac{phem} \ac{wsi} dataset. \ac{phem} describes repeated bleeding into the lung and can have a broad range of causes like, congestive heart failure, leukaemia, physical exercise or autoimmune disorders~\cite{jonckheer2017diffuse,golde1975occult,hopkins1997intense,hinchcliff2005association,martinez2017diffuse,hooi2019bronchoalveolar} with possible life-threatening consequences~\cite{ahmad2019morbidity}. In sport horses a specific disease entity called \ac{eiph} has very high incidences and may lead to reduced athletic performance~\cite{morley2015exercise,hinchcliff2005association,birks2003exercise}. This disease has therefore high relevance for the equine sport industry and has been used as an animal model for human \ac{phem}~\cite{bullone2015asthma}. 
\ac{phem} is often diagnosed by cytologic examination of pulmonary fluid (\ac{balf}) with quantification of hemosiderin content in alveolar macrophages ~\cite{golde1975occult,hoffman2008bronchoalveolar}. In chronic bleeding, macrophages (hemosiderophages) degrade red blood cells into hemosiderin, which is a protein complex containing iron. Usually special stains for iron (such as Prussian Blue or Turnbull's Blue) are used to highlight the hemosiderin concentration in alveolar macrophages. For diagnosis of \ac{phem} in humans a 5-tier grading system has been developed by Golde~\etal~\cite{golde1975occult} and Doucet and Viel~\cite{doucet2002alveolar} have adapted this system for \ac{eiph} in horses. Hooi~\etal~\cite{hooi2019bronchoalveolar} have recently described a similar scoring system for cats.  

For the creation of this novel dataset, we digitised and fully annotated 55 equine, seven feline and 12 human \ac{balf} samples with a total of 297,383 manually verified macrophage annotations in form of bounding boxes. To improve labelling efficiency and data quality, we applied expert-algorithm cooperation in the following manner. Firstly, we incorporated a publicly available pre-trained \ac{eiph} model \cite{marzahl2020deep} for equine \ac{wsi} grading to our multi-species dataset resulting in 585,600 candidate annotations. Secondly, visualisation and clustering techniques were applied to semi-automatically remove 45,944 false positive annotations. Afterwards, a trained pathologist (\acs{cab}) performed a screening and reviewed the complete dataset which left 303,289 hemosiderophages. As a final validation step, the hemosiderophages were arranged and presented according to their grade and conclusively checked by the same trained pathologist (\acs{cab}) resulting in a total of 297,383 manually verified annotated hemosiderophages.     

\begin{figure*}[tbp]
	\centering
	\includegraphics[width=0.73\textwidth]{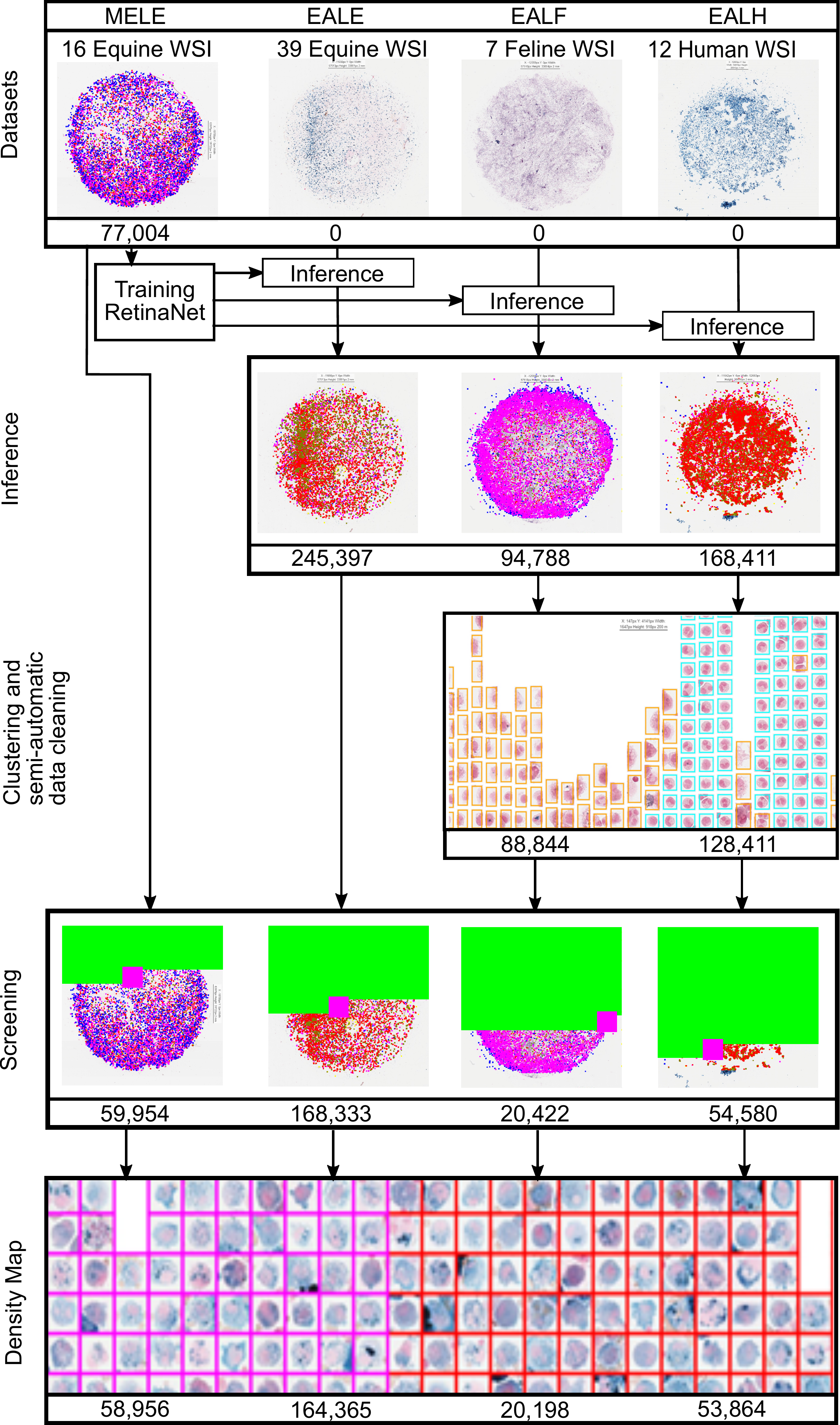}
	\caption{Overview of the macrophage annotation and validation pipeline: The RetinaNet object-detection model trained on 16 equine slides\cite{marzahl2020deep} is used to perform inference on the remaining slides, followed by a semi-automatic clustering step which clusters cells by size. Error-prone cells are highlighted and can then be efficiently deleted by a human expert. Afterwards, a human expert screens all \ac{wsi} to increase the dataset consistency. Finally, a regression-based clustering system is applied to support experts searching for misclassifications of the hemosiderin grade.}
	\label{fig:Overview}
\end{figure*}

As a result of this expert-algorithm pipeline which is visualised in Fig.  \ref{fig:Overview}), we present the largest publicly available fully-annotated multi-species cytology \ac{wsi} dataset to date. Our dataset provides researchers with unprecedented opportunities to develop new inter-species algorithms and can help to overcome domain adaptation limitations. We evaluated the quality of the dataset by conducting a species-wise 3$\times$3 cross validation and performed an ablation study to estimate how many annotated \acp{wsi} are needed to adapt to new species.     

\section{Methods}

The following section describes the sample collection, staining and digitisation procedure followed by our annotation processing pipeline. The \ac{balf} samples of the three species were collected at different institutes for routine diagnostic evaluation of respiratory disease. Therefore, no animal was harmed for the creation of this dataset. Individual case histories were not considered in the present study and all samples we received were anonymised by the providing laboratory. Approval for use of animal specimens was given by the State Office of Health and Social Affairs of Berlin (approval ID: StN 011/20) and for human samples by the University of Lübeck (approval ID: AZ 19-428).
The 74 cytological slides were prepared by cytocentrifugation and stained for iron content with Prussian Blue (n = 37) or modified Turnbull’s Blue using the Quincke reaction (n = 37). Both staining methods result in similar insoluble blue pigments \cite{meguro2007nonheme} and therefore similar hemosiderophages appearances. Digitisation of the glass slide was performed using a linear scanner (Aperio ScanScope CS2, Leica Biosystems, Germany) at a magnification of 400$\times$ (resolution: $ 0.25 ~\frac{\mu m}{px})$. To be as consistent as possible in the data pre-processing phase, all samples were stained and digitised in the same laboratory (Institute of Veterinary Pathology, FU Berlin).

\subsection{Equine datasets}

Fifty-seven equine samples were prospectively collected at the VetPath Laboratory Services (Australia) from 29 \acp{balf} samples of 25 horses with clinical signs of lower respiratory tract disease. Samples were prepared by cytocentrifugation (CYTOPRO 7620, Wescor Inc, Logan, UT, USA) at 510 x g for 3 minutes using a variable volume of \acp{balf} depending on cellular density. Subsequently unstained slides were shipped to the FU Berlin, Germany, and stained with both staining methods and digitalized as described above.   

\subsubsection*{Manually expert labelled equine (MELE) dataset} \label{sec:MELE}

A preliminary dataset using 17 equine \acp{wsi} was developed for a previous publication \cite{marzahl2020deep} and revised for this publication. Initially, these slides were fully annotated by one expert (\acs{cab}) with the open source software SlideRunner \cite{aubreville2018sliderunner} in a two stage process. First all macrophages / hemosiderophages were annotated by screening the \acp{wsi} and afterwards cell annotations were assigned a corresponding grade. From these 17 \acp{wsi}, 16 were added to this publication and one was removed due to a significant fungal contamination (>1\% of the cells) in the Turnbull’s blue staining, resulting in 10 Prussian Blue and 6 Turnbull’s Blue samples from 16 horses. Subsequently (for this publication), the same expert (\acs{cab}) modified this dataset by a second screening process and review of the grades with the help of density maps (see section \ref{sec:DensityMap}). In the following, we will refer to this dataset as \ac{mele} dataset. 

\subsubsection*{Expert-algorithm labelled equine (EALE) dataset} \label{sec:EALE}

For the creation of the \ac{eale} dataset, we used 39 additional \acp{wsi} from 26 horses. A detailed overview regarding the dataset's meta-data can be accessed at the supplementary Table images\_meta\_data.csv. The samples were prepared at the same laboratory as the \ac{mele} dataset and were processed according to the same protocol. The dataset consists of 18 Prussian Blue and 21 Turnbull’s Blue samples. The database was created by interference of the \acp{wsi} with an algorithm developed on the initial dataset (\ref{sec:MELE}) and multiple steps of quality control (\ref{sec:Cluster}, \ref{sec:Screening}, \ref{sec:DensityMap}) as summarised in Fig. \ref{fig:Overview}. 


\subsection{Expert-algorithm labelled feline (EALF) dataset} \label{sec:EALF}

Seven feline samples were retrospectively obtained from the study by Hooi~\etal~\cite{hooi2019bronchoalveolar}, which was designed to evaluate the presence of hemosiderophages in feline \ac{balf} samples. Samples were initially prepared by cytocentrifugation and stained with Wright's stain \cite{hooi2019bronchoalveolar}. For this study specimens were de-stained and re-stained with Turnbull’s Blue. Labels were created by interference and a multi-step quality control (\ref{sec:Cluster}, \ref{sec:Screening}, \ref{sec:DensityMap}). In the following, we will refer to this dataset as \ac{ealf}. 

\subsection{Expert-algorithm labelled human (EALH) dataset} \label{sec:EALH}

The samples were collected by a \ac{balf} procedure using local anaesthesia bronchoscopy. In all cases humans did not undergo any steroid or other immunoregulatory therapy. After the volume of recovered \ac{balf} had been assessed, the fluid was filtered through a layer of sterile gauze, centrifuged (15 min, 4$^\circ$C, 65 x g) and resuspended. Total cell counts were assessed in a Neubauer chamber and viability was determined by trypan blue exclusion. Each cytospin slide was prepared from \ac{balf} with 50,000 cells (600 cpm, 15 min; Heraeus Sepatech Omnifuge 2.0 RS, Hanau, Germany). Following staining with May-Grünwald-Giemsa and HEMATOGNOST Fe® SIGMA routine cytological examination were performed to confirm \ac{phem} due to different underlying diseases. Supplementary preparations were made of 12 cases with proven \ac{phem} and unstained specimens were subsequently send to FU Berlin and three stained with Turnbull’s blue and nine with Prussian Blue. In the following, we will refer to this dataset as \ac{ealh} dataset.

\subsection{Labelling and visualisation platform}

To create this multi-species \ac{wsi} dataset, we used the open source online platform EXACT~\cite{marzahl2020exact}, which was specifically modified for this project. The software supports the creation of this dataset with multiple features which we will briefly summarise in the following section. Manual \ac{wsi} annotations are supported by a special screening mode, which allows for systematic screening of slides in a user-defined magnification while saving the progress per expert and therefore allowing to conveniently resume the work at a later point in time. Furthermore, a bounding box annotation process is streamlined by a single-click annotation mode which incorporates the average hemosiderophages size and therefore minimises the need to further adjust the bounding box to the cell size. 
Annotation versioning supports the tracking of changes and provides detailed and reproducible insights into the development process of datasets. 

\subsection{Inter-species inference from a pre-trained model} \label{sec:Inference}

At the time of dataset development, no annotations for feline or human \ac{phem} slides were publicly available, which resulted in limited options to perform transfer learning-based methods. Therefore, we directly applied the publicly available \cite{marzahl2020deep} equine \ac{phem} deep learning model trained on the \acs{mele} dataset to the \acp{wsi} of the \ac{ealh}, \ac{ealf} and \ac{eale} dataset (Fig. \ref{fig:Overview} Inference). The equine deep learning model uses a custom RetinaNet-model\cite{Lin2017ICCV,marzahl2020deep} optimised for hemosiderophage \ac{wsi} detection. The model was trained with the Adam optimiser on 14 fully annotated \ac{wsi} from the \ac{mele} dataset until convergence was reached by a maximal learning rate schedule of 0.01. The model was validated on three remaining fully annotated \acp{wsi} from the same dataset. 

Inference on the 58 unannotated \acp{wsi} of the \ac{eale}, \ac{ealf}, \ac{ealh} datasets took on average 120 seconds per \ac{wsi} on an NVIDIA Quadro P5000 graphics card. To minimise the probability of missing hemosiderophages, we applied a classification probability threshold of 0.35 to all slides to obtain a highly sensitive and less specific model resulting in 585,600 macrophage / hemosiderophage candidate annotations.

\subsection{Semi-automatic data cleaning via customised clustering} \label{sec:Cluster}

The accuracy of deep learning models depends on multiple factors, which are oftentimes difficult to control. One influencing factor, that may lead to varying results, is the quality of the source dataset, which, in turn, strongly depends on various pre-analytic steps such as image acquisition. Additionally, the label quality used for training deep learning models has a strong influence on the final performance, and for \ac{phem} grading a high inter- and intra-observer variability has previously been described  \cite{10.1007/978-3-030-59710-8_3,marzahl2020deep}. Special stains for iron are ideal to quantify the intracytoplasmatic hemosiderin content (stained as blue pigment), but introduces considerable difficulties in differentiation of different cell types due to the weak staining of cellular components. One additional aspect is the domain shift between species, which might manifests in altered cell morphology and texture compared to the source domain (i.e., equine tissue). An example for this domain shift artefacts is the reduced performance of the initial algorithm on the feline samples due to false-positive detections of granulocytes or multiple bounding box predictions per cell.

To minimise the effect of the above-described implications on this dataset, we established the following semi-automatic pipeline. Firstly, all cell patches of a slide were copied into a new image on the EXACT server and sorted by width in ascending order on the x-axis (Fig. \ref{fig:Overview} 3rd row, Clustering). Predictions were grouped by width to high ratio of the bounding box in a annotation map. Thereby a human expert could remove obvious false positive predictions (small cell types and non-maximum-suppression artefacts) using the web interface. This is implemented by drawing a rectangle with a computer mouse around groups of cells to delete them from the dataset. Aforementioned size-based visualisation also allowed the efficient re-labelling of false-positive granulocytes in feline samples due to their significant smaller cell size compared to macrophages. The semi-automatic data cleaning step removed 17.45\% of the cells created at the inference step. 

\subsection{Experts screening}  \label{sec:Screening}

For labelling data, expert-algorithm collaboration is considered suitable for creating high quality datasets~\cite{10.1007/978-3-030-59710-8_3}. Diligent expert review of algorithmic predictions is indispensable, especially for \acp{wsi} that may potentially exhibit a significant domain shift to the initial training data. To keep the screening process as consistent as possible, the same veterinary pathologist (\acs{cab}) performed all annotation tasks. To enable an efficient validation of all algorithm-created annotations across the \acp{wsi}, we used the screening mode provided by the EXACT software. With this mode, it is possible to check a \ac{wsi} patch by patch and correct errors on a user-selected magnification. An overlap of 15\% per patch is applied, and the expert's progress is saved automatically (Fig. \ref{fig:Overview} Screening). In this screening step, the expert removed 44.8\% of the automatically detected cells (236,367) and introduced 560 new hemosiderophages on 51,110 patches. These numbers are in line with the high sensitivity and low specificity expected from setting low cutoff values for algorithmic predictions. Similar to the screening of the computer aided annotations, the original manual annotations from the equine \ac{mele} dataset initially developed for a previous publication \cite{marzahl2020deep} were reviewed. Here, the expert deleted 17,050 of the 77,004 annotations (22.1\%) and introduced 30 new annotations. Deletion of such a high number of manual and algorithmic labels was mostly attributed to the difficulty of classifying different cell types (macrophages versus other cell types) with the special iron stain. Clear identification of macrophages (including hemosiderophages) in \ac{balf} is largely based on morphology of the cell nucleus which is, however, only very weakly highlighted with iron stains. Cellular size and shape alone are only vague cell classification criteria. We have notices that the task of distinguishing hemosiderophages against neutrophils may be complicated by the positive iron staining of both cell types. While the initial manual labelling of the \ac{mele} had a high sensitivity for labelling hemosiderophages, its re-evaluation suggested that many neutrophils had been wrongly annotated. During expert screening, unambiguous non-macrophagic cells, especially cells with a small cell size, were deleted, however this had no influence on the overall hemosiderin score of the respective \ac{wsi}. 

\subsection{Density map}  \label{sec:DensityMap}

\begin{figure*}[tbp]
	\centering
	\includegraphics[width=0.99\textwidth]{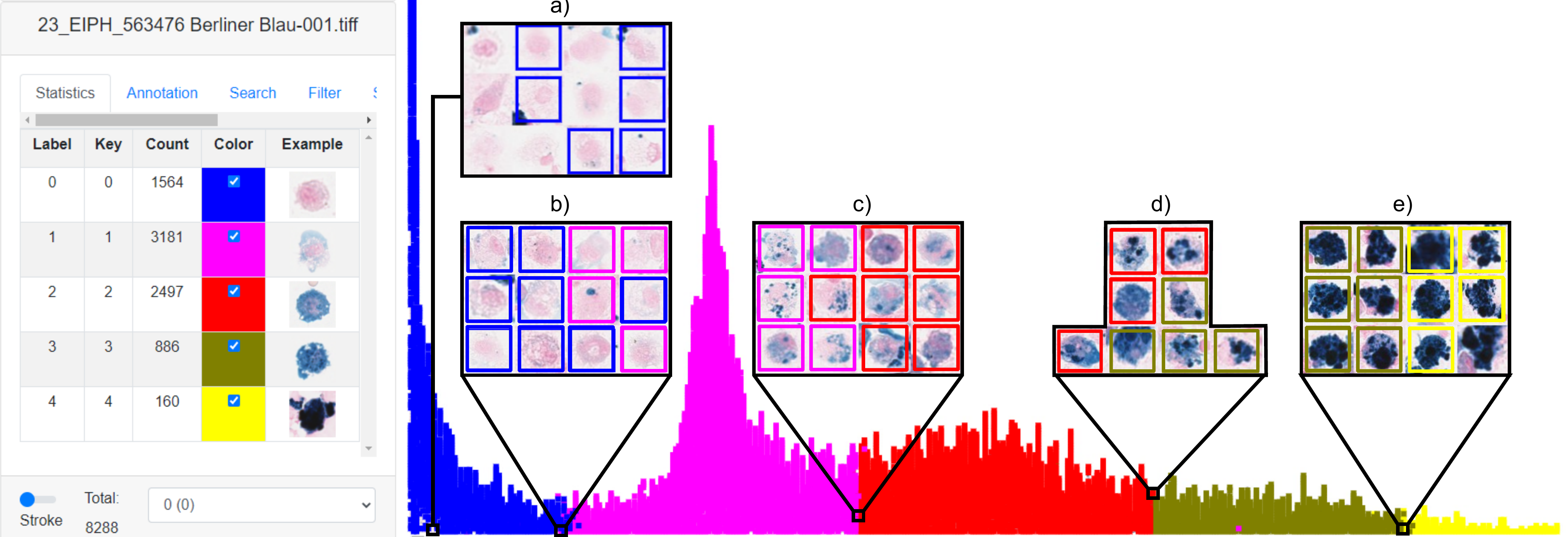}
	\caption{Left: Statistics on the density map at the EXACT user interface. Right: Visualisation of a density map which was screened by an expert for mislabelled cells (especially regarding the label class). Subfigure a) displays six manually deleted annotations. Visualisations b) to e) show the border region between two grades, with the first two columns representing the lower grade and the last two the upper grade, which were sometimes corrected by the expert as visualized by the different color of the bounding box. }
	\label{fig:DensityMap}
\end{figure*}

Initially, all hemosiderophages were classified into discrete grades from zero to four depending on their hemosiderin concentration, both for computer-aided annotations for the \ac{eale}, \ac{ealf}, \ac{ealh} dataset and the expert-created annotations for the \ac{mele} dataset. However, the hemosiderophages hemosiderin absorption is a continuous process which is only mapped to a discrete grading system. This can leads to inconsistent classification between neighbouring grades as previously described by Marzahl~\etal \cite{marzahl2020deep}. To overcome this limitation, we utilised the provided cell-based regression approach \cite{marzahl2020deep} to assign a continuous grade between zero and four to each hemosiderophage. Afterwards, we created a new image-map where the hemosiderophages were arranged in an ascending order along the x-axis according to their hemosiderin score. These novel image-maps were created for each \ac{wsi} individually and reviewed by the same trained pathologist (\acs{cab}) to make the process of identifying mislabelled cells on the border between two grades (Fig. \ref{fig:DensityMap} Density Map) as consistent as possible. On the density maps the expert changed the grade of 38,799 (13.04\% Up: 13,591 Down: 25,208) annotations from which 99.92\% are changes within one grade. The density maps also provided another opportunity to review the cell type of the annotations, which were deleted in 5,906 (1.95\%) instances. 

\section{Data Records}

We provide the 55 equine, 12 human and seven feline original \acp{wsi} in the Aperio SVS format without any identification properties. Alongside, we supply all hemosiderophage annotations after each of the four \nameref{sec:Inference} (Inference), \nameref{sec:Cluster} (Cluster), \nameref{sec:Screening} (Screening), and \nameref{sec:DensityMap} (DensityMap) processing steps as comma-separated files for easy access, as binary files which are compatible to our training and evaluation pipelines and in the sqlite format for SlideRunner \cite{aubreville2018sliderunner}.
Each annotation provides the following information: 
\begin{itemize}
    \item The annotation source slide name
    \item A \ac{uuid}
    \item The absolute bounding box coordinates (x1,y1,x2,y2) on the \ac{wsi} 
    \item The \ac{eiph} grade in a discrete range from zero to four
\end{itemize}

Additionally, we provide a docker build with all packages installed to download the \acp{wsi} and annotations for reproducing our experiments. Table \ref{tab:OVERVIEW} gives an overview of the dataset's meta-data. A detailed per-image statistic can be examined in the supplementary Table images\_meta\_data.csv. The dataset column distinguishes between \ac{mele}, \ac{eale}, \ac{ealf} and \ac{ealh} datasets. The version column indicates the processing steps \nameref{sec:Inference} (Inference), \nameref{sec:Cluster} (Cluster), \nameref{sec:Screening} (Screening), and \nameref{sec:DensityMap} (DensityMap) to which the following statistical data refer. The \ac{eiph} score was calculated by the method of Doucet and Viel~\cite{doucet2002alveolar}.


\begin{table}[]
    \centering
    \begin{tabular}{c|c|c|c|r|c|r|r|r|r|r}
    \toprule
        species & dataset & slides & Version & Total & Score & \multicolumn{5}{c}{Count of Cells by Grade} \\
        &       &        &                                 & Cells &         & 0        & 1        & 2        & 3        &  4    \\ 
     \midrule
        Equine 
               &  \ac{mele}   &  16  &  \ac{mele}  &  77,004         &  102            &  29,017  &  26,810  &  13,178  &  6,577   &  1,422 \\ 
               &         &      &  S     &  59,954         &  112            &  19,733  &  21,545  &  11,442  &  5,963   &  1,271 \\  
               &         &      &  D     &  58,956         &  109            &  19,246  &  21,595  &  11,829  &  5,552   &  734 \\  \cline{2-11}
        
               &  \ac{eale}   &  39  &  I     &  245,397        &  95             &  97,904  &  80,715  &  47,789  &  17,437  &  1,552  \\
               &         &      &  S     &  168,333        &  108            &  54,432  &  60,189  &  39,316  &  13,404  &  992  \\
               &         &      &  D     &  164,365        &  101            &  51,797  &  67,798  &  36,339  &  7,810   &  621 \\ \hline

        Human  &  \ac{ealh}   &  12  &  I     &  168,411        &  133            &  31,035  &  64,833  &  58,320  &  12,776  &  1,447  \\
               &         &      &  C     &  128,012        &  133            &  21,532  &  53,704  &  42,553  &  8,932   &  1,291  \\
               &         &      &  S     &  54,580         &  156            &  47,26   &  20,688  &  23,323  &  5,090   &  753  \\
               &         &      &  D     &  53,864         &  156            &  43,84   &  18,357  &  26,563  &  4,433   &  127  \\ \hline
               
        Feline &  \ac{ealf}   &   7  &  I     &  94,788         &  38             &  58,879  &  35,659  &  122     &  8      &   120  \\
               &         &      &  C     &  88,848         &  38             &  54,867  &  33,868  &  103     &  5      &   5  \\
               &         &      &  S     &  20,422         &  33             &  13,631  &  6,747   &  41      &  2      &   1  \\
               &         &      &  D     &  20,198         &  45             &  11,124  &  9,039   &  35      &  0      &   0  \\ \hline \hline  
               
        Total  &  SDATA  &  74  &  I     &  585,600         &  98            &  216,835 &  208,017 &  119,409 &  36,798  &  4,541 \\
               &         &      &  S     &  303,289         &  113           &  92,522  &  109,169 &  74,122  &  24,459  &  3,017 \\
               &         &      &  D     &  297,383         &  109           &  86,551  &  116,789 &  74,766  &  17,795  &  1,482  \\
\bottomrule
\end{tabular}
    \caption{Overview of the dataset meta-data, including the species, the dataset name, the number of slides, the version of the post-processing refinement step (\textbf{I}nference, \textbf{C}luster, \textbf{S}creening, \textbf{D}ensityMap) and the number of labels per hemosiderin score. }
    \label{tab:OVERVIEW}
\end{table}

In total, the expert screened 51,110 patches on 74 \acp{wsi} from three species which covers a total area of $5,196.17mm^2$. This resulted in 297,383 annotated macrophages / hemosiderophages, making this the largest published multi-species dataset of macrophages / hemosiderophages and one of the largest cytology \ac{wsi} datasets in general.   

\section{Technical Validation}

To gain a deeper understanding of the data and to establish a baseline for future studies, we conducted multiple experiments. Firstly, during the screening phase, we noticed that the expert (\acs{cab}) deleted a high number of his own manually created annotations from the dataset of our previous work (\ac{mele} dataset). Furthermore, our deep learning method, which was trained on these initial annotations, also introduced many false positive annotations even at conservative thresholds. This effect was amplified by the decision to configure the model with a relatively high sensitivity in order to miss as few cells as possible. The observation that the initial object detection model was configured to have high sensitivity (and therefore low specificity) is backed by the statistics that only 560 new hemosiderophages were introduced in the screening phase of the dataset development (\ac{ealh}, \ac{ealf}, \ac{eale}) compared to 229,054 deleted cells. The combination of these effects caused the manual deletion of large quantities of annotations as shown in Table \ref{tab:OVERVIEW}.   
To quantify and compensate for this effect, in the following first experiment, we investigated if the trained deep learning model can be efficiently adapted to this change in annotation behaviour by retraining on the updated annotations from the \ac{mele} dataset created for this publication. In a second experiment, we evaluated inter-species domain transfer and performed an inter-species cross-validation study. This experiment is followed by an ablation study to estimate the quantities of annotations needed to train an accurate \ac{eiph} object detector. 
To evaluate the object detection performance of the models trained in our experiments, we used the \ac{map} metric introduced in the 2007 PASCAL VOC challenge \cite{everingham2010pascal}. 

\subsection{Reevaluation of the inference step}

To investigate whether and how efficient the deep learning model can adapt to the changed annotation behaviour, we trained models with the initial and reviewed \ac{mele} dataset and optimised thresholds for the different species individually. We applied the customised RetinaNet architecture with a ResNet-18 pre-trained on ImageNet. The network was trained with the Adam optimiser using a maximal learning rate of 0.001 until the validation loss started to increase. As a metric to quantify how effective the deep-learning model adapted to the new annotations we calculated the \ac{map} score with an \ac{iou} >0.5 and compared total cell numbers. 
The \ac{map} score increased with the new annotations by 5 percent from 0.66 to 0.71 compared to the object detection results reported in earlier works. This indicates that the experts annotations are more consistent. The optimal threshold calculated on the validation set for equine samples increased from 0.35 to 0.65 and for humane and feline slides from 0.35 to 0.80. The total number of detections decreased from originally 585,600 to 301,109 (ground truth 297,383) while the number of false negatives increased from 560 to 7,351 according to the final dataset. In conclusion the deep-learning model is able to adapt to new annotation behaviours and a stronger focus on finding optimal thresholds could lead to decreased manual interactions but introduces the risk of overlooking false-negative annotations.

\begin{figure*}
	\centering
	\includegraphics[width=0.75\textwidth]{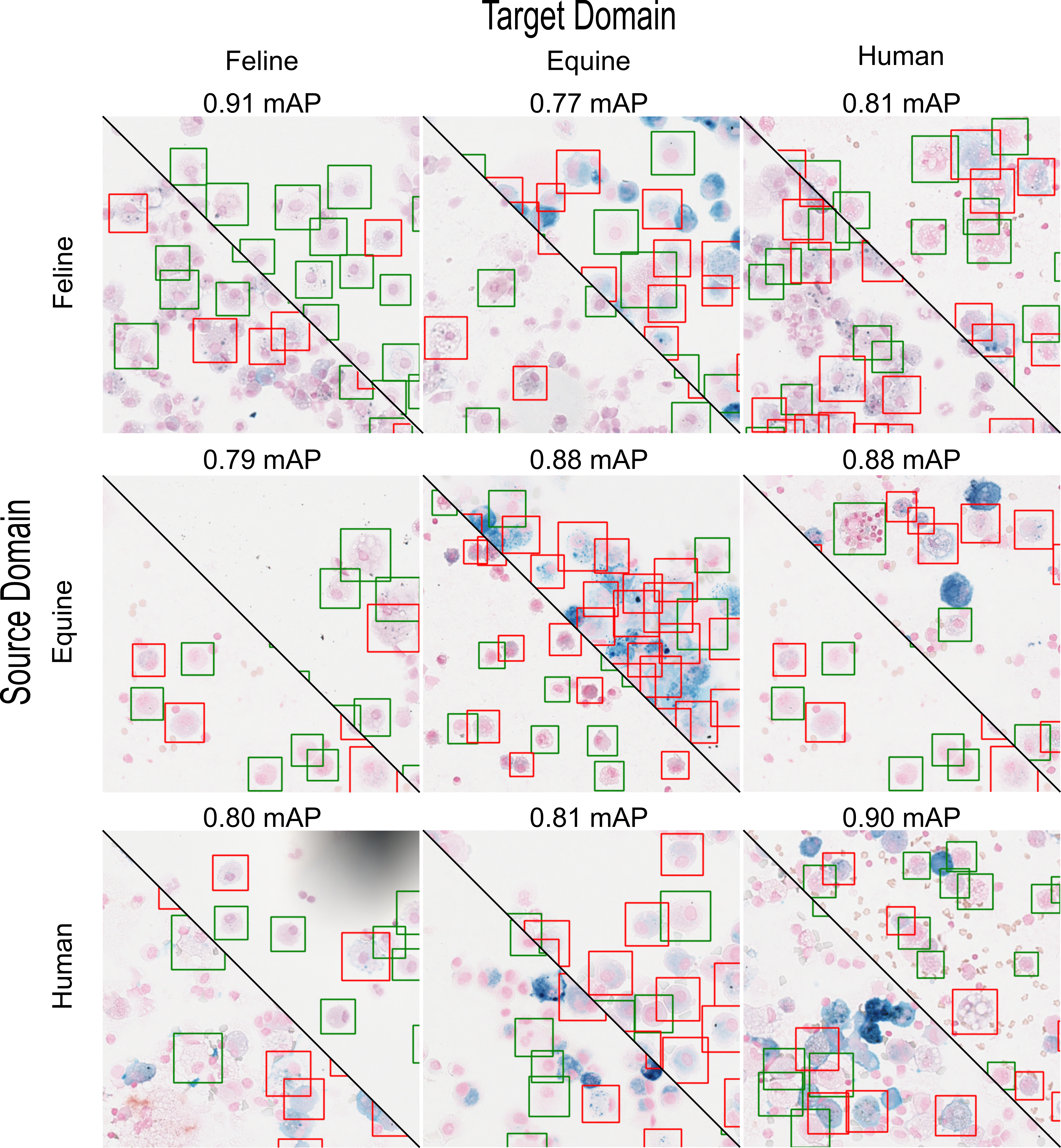}
	\caption{Each of the nine figures show on the left the species source training domain and on the top the species target domain with the obtained \ac{map}. Green bounding boxes represent grade zero hemosiderophages while red show grade one. }
	\label{fig:domainTransfer}
\end{figure*}

\begin{table}[]
    \centering
    \begin{tabular}{l|c|c|c|c|c|c|c|}
    \toprule
        File               & Species    & \multicolumn{5}{c}{Count of Cells by Grade} \\ 
                           &                       & 0    & 1    & 2    & 3     & 4 \\ 
     \midrule

    15.svs		& 	 Equine		& 	 2884	& 	 3124	& 	 2138	& 	 733	& 	 12		\\ 
    22.svs		& 	 Equine		& 	 2147	& 	 2441	& 	 2025	& 	 899	& 	 69		\\ 
    30.svs		& 	 Equine		& 	 2815	& 	 2464	& 	 1523	& 	 606	& 	 26		\\ 
    19.svs		& 	 Equine		& 	 1197	& 	 1012	& 	 316	& 	 36		& 	 0		\\ 
    02.svs		& 	 Equine		& 	 2094	& 	 2560	& 	 1291	& 	 364	& 	 2		\\ \hline  

    2707.svs	& 	 Human		& 	 1434	& 	 2954	& 	 1034	& 	 122	& 	 0		\\ 
    11480.svs	& 	 Human		& 	 391	& 	 1221	& 	 674	& 	 20		& 	 0		\\ 
    10080.svs	& 	 Human		& 	 284	& 	 3015	& 	 2375	& 	 342	& 	 9		\\ 
    10052.svs	& 	 Human		& 	 120	& 	 1739	& 	 3744	& 	 739	& 	 9		\\ 
    10120.svs	& 	 Human		& 	 112	& 	 2225	& 	 4656	& 	 1502	& 	 105	\\ \hline   
    
    1.svs		& 	 Feline 		& 	 1200	& 	 2287	& 	 22		& 	 1		& 	 0		\\ 
    6.svs		& 	 Feline 		& 	 3009	& 	 895	& 	 0		& 	 0		& 	 0		\\ 
    14.svs		& 	 Feline 		& 	 2393	& 	 495	& 	 2		& 	 0		& 	 0		\\ 
    13.svs		& 	 Feline 		& 	 4663	& 	 810	& 	 2		& 	 0		& 	 0		\\ 
    2.svs		& 	 Feline 		& 	 57		& 	 502	& 	 10		& 	 0		& 	 0		\\ \hline \hline 

    27.svs		& 	 Equine		& 	 1392	& 	 1359	& 	 364	& 	 93		& 	 1		\\ 
    17.svs		& 	 Equine		& 	 2725	& 	 2625	& 	 715	& 	 166	& 	 14		\\ \hline  

    10227.svs	& 	 Human		& 	 1265	& 	 1943	& 	 324	& 	 5		& 	 0		\\ 
    2702.svs	& 	 Human		& 	 963	& 	 523	& 	 413	& 	 67		& 	 2		\\ \hline  

    10.svs		& 	 Feline 		& 	 412	& 	 371	& 	 4		& 	 1		& 	 0		\\ 
    12.svs		& 	 Feline 		& 	 1897	& 	 1387	& 	 1		& 	 0		& 	 0		\\ 

    \bottomrule
    \end{tabular}
        \caption{The filenames of the five training and two validation slides (below the double line) per species used for the ablation and inter species cross-validation study. The slides have been selected and ordered according to their ratio of grade zero and one cells to represent a balanced sub-dataset.}
        \label{tab:AblationSlides}
\end{table}

\subsection{Inter-species domain adaptation} \label{sec:domainAdaptation}

As shown by Bullone~\etal \cite{bullone2015asthma}, equines can be used to better understand human asthma on an immunopathological level. To support scientific research in this direction, the use of machine learning models across species is of great scientific and economic importance. To investigate the potential and limitations of transferability across different species, we have carried out a 3$\times$3 cross-validation in which we trained on one species and validated on all other species separately. To support the comparability of the results across the species with their varying amount of available \acp{wsi} and to keep the computational effort within reasonable limits, we decided to use only five \acp{wsi} for training and two other \acp{wsi} for validation (See Table \ref{tab:AblationSlides}). This is further motivated by availability of only 7 feline \acp{wsi}. For the other two species, the training and validation subset was selected by using the seven most balanced slides with respect to the number of grade zero and one macrophages / hemosiderophages (see Table \ref{tab:AblationSlides}). We used this \ac{wsi} sampling strategy to minimise the effect of an imbalanced dataset which could negatively impact the transferability study. Due to the circumstance that feline \acp{wsi} only contain hemosiderophages with the grades zero and one we only used these two classes for the cross-validation for all species and reason that the transferability of these two classes can be generalised to the remaining classes. Example patches and results from this cross-validation experiment are visualised in Figure \ref{fig:domainTransfer}. The experiment achieves best results if the source is equal to the target domain with an \ac{map} value of 0.90 (Equine 0.88, Human 0.90, Feline 0.91). The training on equine slides resulted in an \ac{map} of 0.88 on human data which indicates that a domain transfer without adaptions to the deep learning model might be possible. Further studies need to show if this algorithms can be used for specific disease of humans such as COVID-19~\cite{Drak_Alsibai_2020}. When the source domain is human or feline, the average inter-species \ac{map} is 0.8 (min 0.77, max 0.81).

\subsection{Ablation study} \label{sec:ablationStudy}
\begin{figure*}
	\centering
	\includegraphics[width=0.99\textwidth]{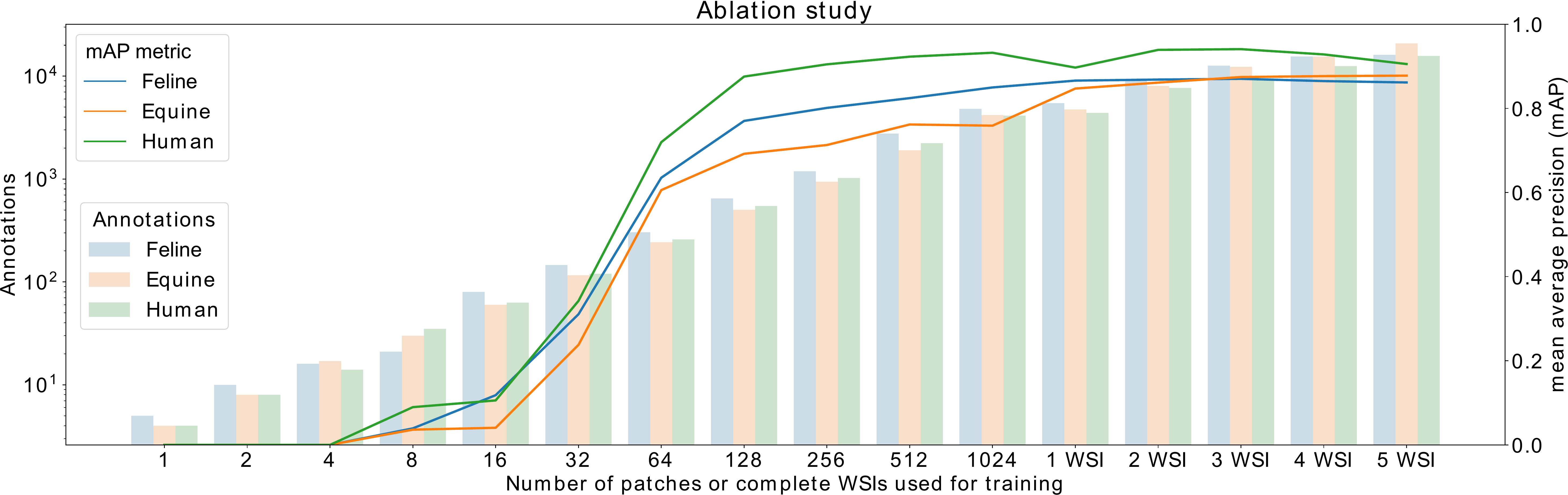}
	\caption{Results of the ablation study using our customised RetinaNet object detector on an increasing number of humane, equine and feline  training patches of size 1024 x 1024 pixel from one \ac{wsi} or up to five complete \acp{wsi}. The boxes represent the total number of hemosiderophages used for training in combination with the \ac{map} graphs for each species.}
	\label{fig:AblationsStudy}
\end{figure*}

Annotating \acp{wsi} manually is a laboursome and expensive task. Therefore, one of the most interesting questions in creating datasets and training deep learning models is the number of \acp{wsi} and annotations needed to reach a converging performance. To answer this question, we started training for each species separately on one uniquely sampled patch (size 1024 x 1024 pixel, number of annotations: mean=6.19, SD=3.74) from one slide and then doubled the number of patches from the same slide every time training reached convergences on the validation set. The training set was chosen to have a balanced number of grade zero and one hemosiderophages. The cell-covered area of each \ac{wsi} contains on average 1,000 unique patches, therefore we continued the ablation study using up to five different \acp{wsi} for training after reaching the values of 1024 training patches on the first slide. To increase the comparability between our experiments, we used the same network, parameters, annotations and slides as described in the section \nameref{sec:domainAdaptation}.
As visualised in Fig. \ref{fig:AblationsStudy}, the performance of the model increased significantly independent of the species until 128 patches with around 1000 unique hemosiderophages and started to converge afterwards even if additional \acp{wsi} were introduced and the total number of annotations was increased up to twentyfold. As described above, to keep the experiments between species comparable, we only used grade one and two hemosiderophages and therefore reason that around five hundred cells per type are sufficient to reach convergence. To put this into perspective: 12 human samples contain only 127 grade four hemosiderophages, making the shown inter-species domain transfer highly valuable for creating deep learning models for human data. This is especially valid for \ac{phem}, which has a particularly high incidence in horses.

\subsection{Code availability and usage notes}

All code used in the experiments to generate results, plots and tables was written in Python and is available through our GitHub repository for \ac{eiph} analysis (\url{https://github.com/ChristianMarzahl/EIPH_WSI/}) in the folder SDATA. A demo-server with all annotations can be accessed at https://exact.cs.fau.de/ with the user "SDATA\_EIPH\_2021" and the password "SDATA\_ALBA". 
The most prominent dependencies are: fast.ai \cite{howard2020fastai}, a deep learning library which is build on PyTorch \cite{paszke2019pytorch}, matplotlib \cite{hunter2007matplotlib} for visualisation, object-detection-fastai with our custom RetinaNet implementation and OpenSlide \cite{goode2013openslide}. The repository is structured as follows: On the top level the "Download.ipynb" jupyter notebook will download all slides and annotations from figshare automatically. The folder \texttt{Statistics} contains notebooks which analyse the dataset annotations and general information about the slides.  
\texttt{Inference} contains code to train the described models and perform inference on slides. 
\texttt{Regression} trains the regression models to predict a continuous \ac{eiph} grade and is used for creating the density maps. 
\texttt{Cluster} contains code to create custom annotation maps and synchronise the generated images and annotations with EXACT.  

The README.md also provides instructions to setup EXACT docker based with all \acp{wsi} and annotations installed and ready to examine.

\bibliography{sample}

\begin{acronym}
\acro{phem}[P-Hem]{Pulmonary hemorrhage}
\acro{wsi}[WSI]{whole slide image}  
\acro{eiph}[EIPH]{exercise-induced pulmonary hemorrhage}  
\acro{balf}[BALF]{bronchoalveolar lavage fluid}  
\acro{cab}[C. A. B.]{Christof~A.~Bertram}
\acro{mele}[MELE]{manually expert labelled equine}
\acro{eale}[EALE]{expert-algorithm labelled equine}
\acro{ealf}[EALF]{expert-algorithm labelled feline}
\acro{ealh}[EALH]{expert-algorithm labelled human}
\acro{uuid}[UUID]{universally unique identifier}
\acro{ap}[AP]{Average Precision}
\acro{map}[mAP]{mean Average Precision}
\acro{iou}[IoU]{intersection over union}

\end{acronym}

\section*{Acknowledgements}

C. A. B. gratefully acknowledges financial support received from the Dres. Jutta \& Georg Bruns-Stiftung für innovative Veterinärmedizin.

Human samples were kindly provided by the BioMaterialBank Nord, which is supported by the German Center of Lung Research. The BioMaterialBank Nord is member of popgen 2.0 network (P2N) which is supported by a grand from the German Ministry for Education and Research (01EY1103). 

\section*{Author contributions statement}

C. M. and C. A. B. wrote the manuscript and carried out the main research and analysis tasks of this work. C. M. carried out data analysis, training of networks and method development. C. A. B. was responsible for image annotations. J. H and J. S. provided the equine, D. B. the feline and L. W. the human samples. M. A., F. W., J. V., K. B., R. K., A. M. provided expertise through intense discussions.

All authors contributed to the preparation of the manuscript and approved of the final manuscript for publication. 

\section*{Competing interests}

The author(s) declare no competing financial interests.
    
\end{document}